\documentclass[acmsmall]{acmart}

\usepackage{booktabs} 
\usepackage{multirow}
\usepackage{tcolorbox}
\usepackage{subfigure}

\usepackage[ruled]{algorithm2e} 

 \newcommand{\tabincell}[2]{
    \begin{tabular} {@{ }#1@{ }}#2\end{tabular}
 }

\SetAlFnt{\small}
\SetAlCapFnt{\small}
\SetAlCapNameFnt{\small}
\SetAlCapHSkip{0pt}
\IncMargin{-\parindent}

\acmJournal{TKDD}
\acmVolume{0}
\acmNumber{0}
\acmArticle{00}
\acmYear{2020}
\acmMonth{0}
\acmArticleSeq{00}

\setcopyright{usgovmixed}

\acmDOI{0000001.0000001}

\begin{document}
\title{The Gene of Scientific Success}

\author{Xiangjie Kong}
\affiliation{%
\department{College of Computer Science and Technology}
 \institution{Zhejiang University of Technology}
\city{Hangzhou}
\postcode{310023}
\state{Zhejiang}
\country{China}
 }
\email{xjkong@acm.org}

\author{Jun Zhang}

\affiliation{%
\department{Graduate School of Education}
 \institution{Dalian University of Technology}
\city{Dalian}
\postcode{116024}
\state{Liaoning}
\country{China}
 }
\email{junzhang@dlut.edu.cn}

\author{Da Zhang}

\affiliation{%
\department{Department of Electrical and Computer Engineering}
  \institution{University of Miami}
\streetaddress{5452 Coral Gables}
\city{Miami}
\state{FL}
\postcode{33124}
\country{USA}
  }
\email{zhang.1855@miami.edu}

\author{Yi Bu}
\affiliation{%
  \institution{Peking University}
\department{Department of Information Management}
\city{Beijing}
\postcode{100871}
\country{China}
}
\email{buyipku@gmail.com}
\author{Ying Ding}
\affiliation{%
  \institution{University of Texas at Austin}
\department{School of Information}
\city{Austin}
\state{TX}
\postcode{78701}
\country{USA}
}
\email{ying.ding@austin.utexas.edu}

\author{Feng Xia}
\authornote{Corresponding author}
\affiliation{%
 \department{School of Science, Engineering and Information Technology}
 \institution{Federation University Australia}
\city{Ballarat}
\state{VIC}
\postcode{3353}
\country{Australia}
 }
\email{f.xia@acm.org}
\thanks{The authors would like to thank
Shenwei Zhang and Wenjie Kang for their helps with the experiments.}

\begin{abstract}
This  paper  elaborates how to identify and evaluate causal factors to improve scientific impact. Currently, analyzing scientific impact can be beneficial to various academic activities including funding application, mentor recommendation, and discovering potential cooperators etc. It is universally acknowledged that high-impact scholars often have more opportunities to receive awards as an encouragement for their hard working. Therefore, scholars spend great efforts in making scientific achievements and improving scientific impact during their academic life. However, what are the determinate factors that control scholars' academic success? The answer to this question can help scholars conduct their research more efficiently. Under this consideration, our paper presents and analyzes the causal factors that are crucial for scholars' academic success. We first propose five major factors including article-centered factors, author-centered factors, venue-centered factors, institution-centered factors, and temporal factors. Then, we apply recent advanced machine learning algorithms and jackknife method to assess the importance of each causal factor. Our empirical results show that author-centered and article-centered factors have the highest relevancy to scholars' future success in the computer science area. Additionally, we discover an interesting phenomenon that the $h$-index of scholars within the same institution or university are actually very close to each other.
\end{abstract}

%
%


%
%



\maketitle

\renewcommand{\shortauthors}{X. Kong et al.}

\section{Introduction}
The development of our society highly associates with scientists' diligent research comments. To recognize and support their contributions, a series of awards and research opportunities are given to these outstanding researchers. Recently, many researchers focus on how to determine the outstanding scholars and how to become a successful scholar, and many indicators from different aspects are proposed to quantify this scientific success \citep{RN4, kong2019the}. However, among these diverse indicators, which factor(s) is decisive and contributive to the academic success remains to be explored. Moreover, causal effects learning is a fundamental problem in machine learning with applications in various fields such as biology, economics, epidemiology, and computer science \citep{spirtes2016causal,wu2018interactive}. Inspired by the above observations, our paper focuses on learning the causal effects that play vital roles in scholars' academic success.

Quantifying the scientific success of scholars has always been an interesting topic that attracts researchers with diverse backgrounds to study \citep{xia2017big, sinatra2016quantifying, Wang2017Scientific, Fortunato2018Science}. The goal of science of success is to first understand the underlying mechanism and then discover a generative model to predict what the values of scientific success would be taken into account outside causal factors. Based on citation counts, a series of evaluation metrics have been proposed, such as journal impact factor \citep{garfield2006history}, g-index \citep{egghe2006theory}, and h-index \citep{hirsch2005index} etc. B. Van Houten \citep{van2000evaluating} defines scientific impact as peer evaluation of scientific research academic works and other achievements, the importance of scientific impact depends on their research achievements being valued, recognized and cited by others. To measure the scientific impact, researchers have identified various controlling factors to capture the diverse characteristics of scholarly entities. Among these factors, citation counts have been regarded as the primary factor to evaluate the scientific impact for its simplicity and efficiency.

Besides the citation-based metrics, scientists also investigate this question from the perspective of network topologies. Initially, the importance of ranking algorithms, such as PageRank \citep{page1999pagerank} and HITS algorithms \citep{Kleinberg:1999:ASH:324133.324140}, are designed for ranking web pages' importance. Recently, inspired by these importance ranking algorithms, researchers also widely utilize them to evaluate the scientific impact in academic networks \citep{amjad2016muice, Dunaiski2016Evaluating}. Considering the merits of both PageRank and HITS algorithms, Wang et al. \citep{Wang2016Coranking} propose the MRCoRank method to measure the impact of scholarly entities in heterogeneous academic networks through mutual reinforcement.

Additionally, the development of social media enables scholars share their articles regularly on Twitter or Facebook. This information sharing effect is no more limited to academic social networks. It has spread out to many digital libraries and is prevalently used across social media platforms. Along with this trend, Altmetrics is proposed as another benchmark to measure the popularity of scholars and their publications by assessing their obtained social attentions. Meanwhile, researchers start utilizing the Altmetrics to quantify the researcher's scientific impact since it can capture the early impact promptly. For instance, Bornmann et al. \citep{Bornmann2016How} normalize publications' Twitter counts to measure the impact of research, and then use it for cross-field comparisons. Furthermore, lots of research methods explore the correlation between Altmetrics and citation counts based methods by statistically analyzing their interrelationship \citep{Feng2016Bibliographic, Costas2015Do}.

The evaluation of scientific impact can shed light on diverse practical issues, such as awards or funding applications, job employments and advisor choosing \citep{liu2019shifu}. Commonly, successful scientific scholars obtain extra opportunity to acquire research resources, receive grant, and spread their research idea more extensively. Therefore, scholars aspire to improve their scientific impact continually. However, what factors have causal relationships with scientific success?  Additionally, which factors are most appropriate to evaluate the scientific success? The above questions still remain unresolved. Therefore, in this paper, we conduct researches on first, identifying causal factors that contribute to the scientific success and then gauge the causality importance of these factors. In this paper, we take the most commonly used h-index as the metric for evaluating scholars' impact and mine the causal factors that lead to scholar's high h-index.

\begin{figure}
\centering
\includegraphics[width=7.5cm]{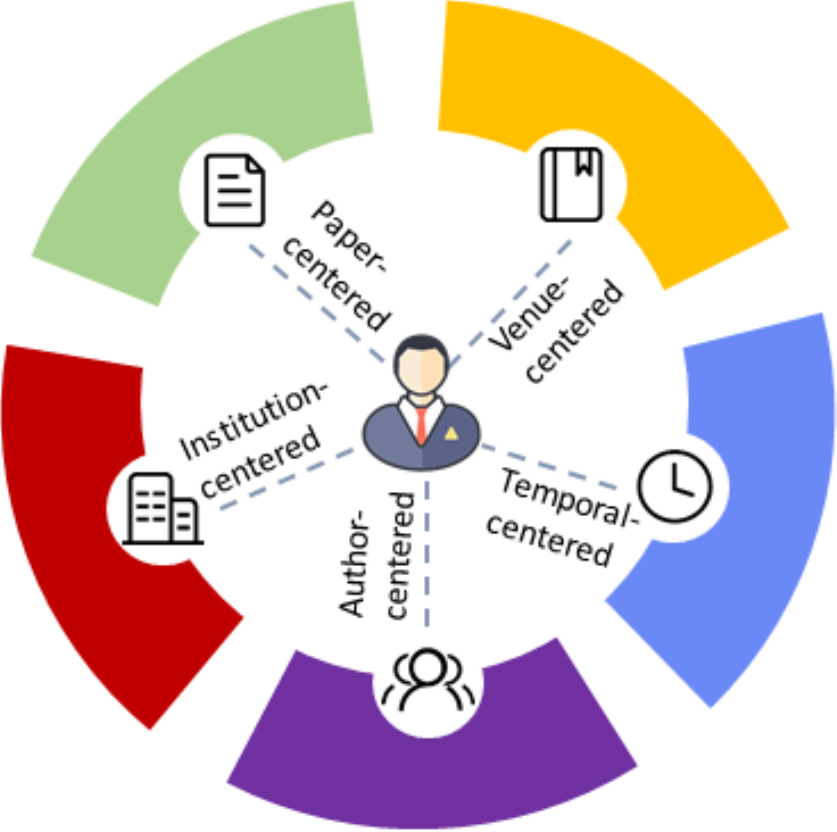}

\caption{Illustration of a scholar's impact relevant factors.}
\label{1}
\end{figure}


Due to privacy issues and technology limitations, the publications' information is easier to access compared to other data sources. It can also represent the corresponding scholars' academic abilities or contributions considerably. Generally, the title, keywords, authors, institutions, venue, pages, and published dates of a publication can be directly obtained. Based on these data, a lot of impact factors can be extracted and calculated as most of the current work does. While unlike previous work, our method does not focus on improving the evaluation metrics, instead we are aiming at discovering the causal factors that affect scholars' academic success from their publications. To answer this question, we categorized the impact factors into several categories as shown in Figure \ref{1}. They are article factor, author factor, venue factor, institution factor, and temporal factor. For each factor, we propose concrete and intuitive indicators to represent each scholar's academic characters. After that, by utilizing the machine learning algorithms and jackknife method, we explore the contribution of each factor on scholars' academic success.

\textbf{Contribution.} Our research mainly focuses on discovering causal factors of scholars' academic success. In general, we make the following contributions in this paper:

\begin{itemize}
\item \textbf{Novel features.} We present five potential causal factors taking the novel Gini Coefficient of institutions into account.
\item \textbf{Causal detection.} Through utilizing the machine learning algorithms and jackknife method, we find that scholars' author-centered and article-centered factors are highly correlated with their academic success.
\item \textbf{New insight.} Our findings provide researchers a novel and efficient method to improve their scientific impact.
\end{itemize}

The rest of the paper is organized as follows. Section 2 discusses related work. Section 3 identifies the proposed scientific impact factors. Section 4 verifies the causal factors, assess their significance and testify them over big scholar dataset. Then, we conclude our work in Section 5.

\section{Related Work}
 The scientific impact has been studied for decades by researchers from a variety of disciplines. For a long time, citation counts have been widely applied to measure the scientific impact. Along with this tendency, researchers have proposed various citation-based metrics. With the developments of academic networks, scholars also look into the scientific impact problem from the angle of network importance. These indicators can be employed for both impact evaluation and future academic success predictions. In this section, we will introduce the related work from the above-mentioned aspects respectively.

The citation count was first utilized to quantify the impact of journals. From then on, researchers have proposed a variety of citation-based methods to measure the scientific impact \citep{valenzuela2015identifying}, such as the h-index \citep{hirsch2005index} and g-index \citep{egghe2006theory}. Some scholars claim that citations should not be regarded as equal \citep{Bai2017The}. Another example that should be mentioned here is: Both A and B cite C. Previously the citations from A and B are regarded as equal, but if A is from a  highly-cited paper while B is not, the citations should be differentiated. This is similar to PageRank-related ideas. A series of approaches for distinguishing the importance of citations have been proposed \citep{valenzuela2015identifying}. These methods all applied the citation counts as an important part of the evaluation metrics, while they all make some improvements since simply relying on citation counts is unilateral for impact evaluation \citep{wang2013quantifying}.

Besides utilizing citation to quantify the scientific impact, scholarly networks are now frequently applied to study such problems since the networks contain various types of entities and relationships \citep{liang2016scientific}. The PageRank and HITS algorithms are widely used for measuring the scientific impact in academic networks. On the basis of these two algorithms, a series of network-based evaluation metrics have been proposed \citep{yu2017multiple}. Considering the effect of different academic network structures, scholars apply the modified importance ranking algorithms to evaluate the impact of different scholarly entities \citep{amjad2015topic}. Other than considering network topologies, some researchers also discover novel features and relationships to evaluate the scientific impact. Wang et al. \citep{Wang2016Coranking} rank the impact of scholarly entities by exploring the text features in heterogeneous academic networks. Due to the evolvement nature of academic networks, some studies also consider the dynamics of citations and the new emergence of new entities or relationships to evaluate the scientific impact \citep{Zhang2017Exploring, amjad2017standing}.

Other than using citation-based and network-based features to evaluate the scientific impact, scholars also try to explore the relevant factors which are very crucial for the future academic performance and predict the future impact \citep{Cao, Heiberger2016Choosing}. Wang et al. \citep{wang2013quantifying} verify the effectiveness of early citations in predicting the potential citations of articles \citep{Klimek2016Successful, Cao2016A}. Stegehuis et al. \citep{Stegehuis2015Predicting} utilize two significant factors, namely historical citation information and journal impact factor to predict papers citation distribution.

The prediction of scholars' future influence, $h$-index, and future citations are all within the scope of future impact prediction \citep{wang2013quantifying, clauset2017data}. Acuna et al. \citep{Acuna2012Future} apply the number of papers, $h$-index, and academic ages of a scholar to predict his/her future impact. The linear regression method is utilized to predict the future impact of outstanding scholars from mathematics, physics and biology research area. And they found that the academic ages of scholars actually play a significant role in predicting scientific impact \citep{Li2015The}. Additionally, Dong et al. \citep{Dong2016Can} study the question of which paper can increase scholar's $h$-index through the linear regression method. They discover that among six factors, the topic and venue are very crucial for the predictions.

Scientific impact prediction with causal inference is a recently emerging research field. Unlike previous methods \citep{wang2013quantifying, clauset2017data, Acuna2012Future}, causal inference based methods first identify potential causal factors and then use them to guide the scientific impact prediction. Additionally, previous methods only consider a single perspective for assessing the scientific impact while neglecting analyzing and ranking the importance of each causal factor.

\section{Causal Factor Identification}
Researchers have studied the problem of scientific impact for decades and propose a variety of impact factors. However, there is no formal definition for scientific impact and no commonly accepted standard for scientific impact evaluation up to now. Among these previously studied impact factors, which factors are most relevant to scholars' academic success? Discovering the answer to it can help researchers carry out their research more efficiently. In this section, we will introduce several novel impact factors, organize the existing factors, and classify them into different categories. 

\subsection{Article-centered Factors}
Generally, most previous work prefers using the citation counts and the number of articles to quantify the scientific impact. While beyond these two indicators, there exist diverse article-based factors that affect the dynamics of scientific impact. To discover the representative features for articles, we first analyze the elements related to article-centered factors.

Citation counts ($Cits$), and the number of publications ($Num_{pub}$) are the basis of article-based factors. The average citations for each scholar ($Ave_{ci}^{a_{i}}$), the highest citations ($Hi_{ci}^{a_{i}}$), the lowest citations ($Lo_{ci}^{a_{i}}$) can be directly obtained through the values of their total $Cites$ and ($Num_{pub}$).
Moreover, the quality of an article depends not only on its content, but also on its topic popularity. For instance, previously, a wide variety of data cannot be acquired and processed due to the technical limitations. While with the developments of data processing technologies and advancement of big data era, paper related to big data topics receive more attention recently. Consequently, the topics of an article also can affect its influence. In order to capture this character, we propose the article's topic popular degree (ATP), which can be calculated according to the following equation.
\begin{equation}
\label{equ:2}
ATP(p_{i}) =\frac{\sum_{w=1}^{m}Num(w)^{p_{i}}}{\sum_{i=1}^{n}Num(i)}
\end{equation}
where $p_{i}$ represents the paper, $w$ is the keyword of paper $p_{i}$, $Num(w)$ is the number of $w$, $m$ is the number of keywords in paper $p_{i}$, $Num(i)$ is the number of the keywords of papers, and $n$ is the total number of publications.

Besides the above mentioned citation-based factors, the qualities of references also need to be considered when measuring the scientific impact. Generally, every scholar has a list of publications, and each publication has a series of references. Citations can be deemed as academic acknowledgments from other researchers. Similarly, the authors of an article also are enlightened by its references. Therefore, references can affect the quality of an article. Primarily, the highest ($Hi_{ci}^{ref}$), the average ($Ave_{ci}^{ref}$), the lowest citation counts ($Lo_{ci}^{ref}$), and the average number of references ($Ave_{num}^{ref}$) are the most direct measurements to quantify the qualities of references. Beyond the citations, the impact of references' venues is also utilized to evaluate the impact of references since many researchers tend to cite articles from high impact venues regardless of the relevance between articles.

To measure the relevance between articles ($Rel_{ref}$), we first solve this problem from the angle of authors. According to each author's publications, their research areas can be represented by exacting articles' keywords. Therefore, we utilize the differences among authors' keywords to calculate the relevance between articles and references. The information entropy is applied to quantify it, and the calculation formula is as follows:
\begin{equation}
\label{equ:1}
Rel_{ref}^{p\rightarrow q} = - \sum_{i=1}^{r}{W}_{i}\log_{2} \left({W}_{i} \right)
\end{equation}
where $Rel_{ref}^{p\rightarrow q}$ represents the relevance between article $q$ and its reference $p$, $W_{i}$ is word's frequency in article $q$ and $p$'s keywords' information, and $r$ is words' total counts.

Furthermore, the relevance between the articles and their reference also needs to be considered. Due to unavailability of articles' full texts, we use the cosine similarity to measure the relevance between the articles' and their references' titles and keywords. For each article and its reference, we extract the sequence of words ($m_{1}, m_{2}, m_{3}, ..., m_{n}$) from their titles and keywords. Then a vector can be obtained based on the above sequence for each paper. According to these vectors, the relevance between an article and its reference can be calculated as follows:
\begin{equation}
\label{equ:2}
Sim(p_{1},p_{2}) = \frac{\sum_{i=1}^{n}(V_{p_{1},i}*V_{p_{2},i})}{\sqrt{\sum_{i=1}^{n}V_{p_{1},i}^{2}}*\sqrt{\sum_{i=1}^{n}V_{p_{2},i}^{2}}}
\end{equation}
where $Sim(p_{1},p_{2})$ represents relevance between paper $p_{1}$ and $p_{2}$, $V_{p_{1}}$ is vector of $p_{1}$, and $V_{p_{2}}$ is $p_{2}$'s vector.

\subsection{Venue-centered Factors}
Besides citation-based metric, PageRank can also be used to measure the qualities of the venues, which reflects the scientific success of an author. To gauge the importance of venues, the PageRank values ($PR(v_{i})$) in the paper-venue network are first calculated. Then, the average citations of papers published in the venues ($Ave_{ci}^{v_{i}}$) is used to measure the quality of venues. Furthermore, with the aids of the concept of scholar's $h$-index, we calculate the $h$-index of venue ($h(v_{i})$). Specifically, the definition of the $h$-index of a venue is similar to the original calculation procedure of $h$-index, and the $h$-index value of a venue equals to $h$ that at least $h$ papers in the venue have $h$ citations.

\begin{equation}
\label{equ:2}
PR(v_{i})=  \sum_{j=1}^{n} Ave_{cj}^{v_{i}}
\end{equation}

\subsection{Author-centered Factors}

Aside from article-centered factors, the factors that represent scholars' attributes are vital to their impact as well. Each scholar's $h$-index ($h_{a_{i}}$) and PageRank value ($PR_{a_{i}}$) in collaboration network are intuitive factors to indicate a scholar's impact. Meanwhile, the journal impact factor (JIF) can be calculated based on them, and is widely applied to measure the impact of journals for its simplicity. According to the concept of JIF, the author's impact factor (AIF) is proposed. Similarly, the AIF of a scholar in year $t$ is scholar's $Ave_{ci}$ in $\triangle t$ years before year $t$. Besides the citation counts, the sum of PageRank scores of scholars' papers $PR_{pub}$ in citation network also can indicate their importance.

Other than these two factors, scholars have proposed several well-known factors to quantify the dynamics of scholars' impact. The $Q$ value is widely applied to reveal the mutual reinforce process of scholars' impact on their papers \cite{sinatra2016quantifying} and is stable during scientists' whole academic careers. The calculation formula of Q value is as follows:
\begin{equation}
\label{equ:1}
Q(a_{i}) = e^{\langle logc_{i\alpha} \rangle}-\mu_{p}
\end{equation}
where $Q(a_{i})$ represents scholar Q value, $\langle logc_{i\alpha} \rangle$ is $a_{i}$'s average citations in logarithmic way, $\alpha$ is $a_{i}$'s $\alpha$-th article, and $\mu_p$ is the average potential influence of articles.

While in each scholar's academic career, they will encounter a variety of researchers from different disciplines. Scholars will benefit from the academic exchanges and discussions with other researchers, and furthermore improve their own scientific impact. As a consequence, the capacities of coauthors also can affect the qualities of their articles and scholars' impact in the meantime. To capture coauthors' influence, several factors are proposed. Typically, the $h$-index of coauthors represents their abilities. Based on it, a series of factors can be easily obtained. The max ($hmax_{co}^{a_{i}}$) and average values ($have_{co}^{a_{i}}$) of scholar $a_{i}$s' coauthors can be directly acquired through each author's $h$-index. Then we use the differentials between $a_{i}$'s $h$-index and $hmax_{co}^{a_{i}}$ ($hdif_{a_{i}}$) to represent the distance between influential coauthors and $a_{i}$.

Besides using the $h$-index to quantify coauthors' academic capacities, we then consider the effects of coauthors' diverse research backgrounds on scholars' impact. Since co-operations among researchers are getting more and more frequently, the integration of scholars from multi-disciplines also has the positive influence on promoting the developments of science and technologies. To measure the range of coauthors' disciplines, we apply the theory of entropy. The detail information on scholars' specific disciplines and institutions can be obtained from the dataset. For each scholar, we quantify the diversity of his or her coauthors ($Div(a_{i})$) by utilizing the theory of entropy. The diversity is computed according to the following equation.
\begin{equation}
\label{equ:1}
Div (a_{i})_{inst} = - \sum_{m=1}^{r}{w}_{m}\log_{2} \left({w}_{m} \right)
\end{equation}
\begin{equation}
\label{equ:2}
Div (a_{i})_{key} = - \sum_{\rho=1}^{q}{k}_{\rho}\log_{2} \left({k}_{\rho} \right)
\end{equation}
\begin{equation}
\label{equ:3}
Div (a_{i}) = Div (a_{i})_{inst} + Div (a_{i})_{key}
\end{equation}
where $Div (a_{i})_{inst}$ and $Div (a_{i})_{key}$ represent author $a_{i}$'s diversity of cooperators' institutions and their papers' keywords, and $Div (a_{i})$ indicates $a_{i}$'s overall cooperators' diversities. $w_{m}$ is word $m$'s frequency in the overall $a_{i}$'s cooperators' institutions' information, and $r$ is word $m$'s total counts in Eq. (\ref{equ:1}). $k_{\rho}$ is word $\rho$'s frequency in all $a_{i}$'s cooperators' papers' keywords, and $q$ is the total number of word $\rho$.

\subsection{Institution-centered Factors}
The effects of institutions on scholars' impact also need to be considered since research funding or policy issues can significantly influence researchers' progress on their studies. Meanwhile, scholars' academic achievements also can be affected by the capacities of their colleagues because they may frequently share research ideas and techniques. Generally, we explore the effects of institutions from two major aspects: scholars' academic environments and the economic factors.

We measure the academic environments from the perspective of colleagues. When conducting researches, people tend to exchange idea with their  co-authors or colleagues. Additionally, researchers are also affected by the influencing group or individuals in their institution. This influence is usually defined as peer pressure (or social pressure).  Taking this peer pressure influence into account, we try to identify the relevance of peer pressure on scholars' academic performance. In other words, is there an actual relationship between them?  To answer the above questions, we proposed several factors to reveal the correlation between scholars' academic success and their colleagues. Initially, we gauge the research capacities of scholars' colleagues. For each scholar, his or her colleague's $h$-index ($h_{col}$), number of publications ($Num_{pub}^{col}$), citation counts ($Cits_{col}$), and PageRank score ($PR_{col}$) can be calculated over the dataset.

Furthermore, we employ the concept of Gini coefficient from the economic field to describe academic reputation of an institution. The Gini coefficient originally utilizes the definition of Lorenz global curves to compute the distribution of income in economic field. Its value ranges from $0$ to $1$. The bigger the value is, the more economic inequality is. In our paper, we quantify the Gini coefficient of institutions using the values of scholar's $h$-index, citations, and the number of papers. The Gini coefficient of an institution can be calculated as follows:
\begin{equation}
\label{equ:1}
G(i) = 1- \frac{1}{n}(2\sum_{m=1}^{n-1}P_{m}+1)
\end{equation}
Here, $G(i)$ represents the Gini coefficient value of institution $i$ and $n$ is the number of research groups within the institution $i$. For a research group, $m$ indicates the group index among $n$ groups. Also, $P_{m}$ is the proportion of the sum of group $m$ in the whole values of institution $i$. Therefore, according to the values of scholar's $h$-index, citations, and the number of papers, the reputation of each institution are calculated using three Gini coefficient values which are $G(i)^{h}$, $G(i)^{Cit}$, and $G(i)^{pub}$.



\subsection{Temporal Factors}
Previous studies have verified the effect of temporal dynamics on scientific impact, such as predicting academic rising stars. For young researchers, they may have a fast growth stage after starting the academic career. The performance during this period is very crucial for their future academic success. We propose two temporal factors to capture this phenomenon. The first one is the academic ages ($Num_{years}$), which are the years since scholars publish their first academic papers. Another  factor is scholars' dynamics of $h$-index during $\triangle t$ years. In this paper, we set the $\triangle t=3, 5, 7$, and then calculate the difference ($Hindex$-$dif$) between the predicted time and $\triangle t$ years ago.

\begin{equation}
\label{pearson}
\rho=\frac{cov(X,Y)}{\sigma X \sigma Y} \\ =\frac{E(XY)-E(X)E(Y)}{\sqrt{E(X^2)-E^2(X)}\sqrt{E(Y^2)-E^2(Y)}}
\end{equation}
where $cov$ is the covariance between two groups of results, and $\sigma$ indicates their standard deviation. Its value ranges from $-1$ to $1$ with correlation varying from the most negative to the most positive.

From the above-mentioned factors, we try to list the relevant indicators of scholars' academic success as comprehensive possible. These indicators are categorized into five major categories, which are article-centered factors, author-centered factors, venue-centered factors, institution-centered factors, and temporal factors. These factors are described in Table \ref{PCC}. Meanwhile, we primarily investigate the correlation between these factors and scholar's $h$-index through the most direct way. The Pearson Correlation Coefficient is applied to measure the relevance between two ranking results. The calculation procedure of Pearson Correlation Coefficient is shown as follows:

\begin{table*}[]
	\centering
	\caption{Causal Factor Descriptions and Correlations.}
	\label{PCC}
	\begin{tabular}{|l|l|l|c|}
		\hline
		& Feature             & Description                                                                              & Correlation \\ \hline
		\multicolumn{1}{|c|}{\multirow{12}{*}{Article}} & $Cits$              & The citation counts of scholars.                                                         & 0.7629      \\ \cline{2-4}
		\multicolumn{1}{|c|}{}                                   & $Num_{pub}$         & The number of publications of scholars.                                                  & 0.7782      \\ \cline{2-4}
		\multicolumn{1}{|c|}{}                                   & $Ave_{ci}$          & The average citations of each scholar.                                                   & 0.2772      \\ \cline{2-4}
		\multicolumn{1}{|c|}{}                                   & $Hi_{ci}$           & The highest citations of each scholar.                                                   & 0.2349      \\ \cline{2-4}
		\multicolumn{1}{|c|}{}                                   & $Lo_{ci}$           & The lowest citations of each scholar.                                                    & 0.2067      \\ \cline{2-4}
		\multicolumn{1}{|c|}{}                                   & $ATP$               & The article's topic popular degree.                                                      & 0.0134      \\ \cline{2-4}
		\multicolumn{1}{|c|}{}                                   & $Hi_{ci}^{ref}$     & The highest citations of references.                                                     & 0.1648      \\ \cline{2-4}
		\multicolumn{1}{|c|}{}                                   & $Ave_{ci}^{ref}$    & The average citations of references.                                                     & 0.1439      \\ \cline{2-4}
		\multicolumn{1}{|c|}{}                                   & $Lo_{ci}^{ref}$     & The lowest citations of references.                                                      & 0.0648      \\ \cline{2-4}
		\multicolumn{1}{|c|}{}                                   & $Ave_{num}^{ref}$   & The average number of references.                                                        & 0.2496      \\ \cline{2-4}
		\multicolumn{1}{|c|}{}                                   & $Rel_{ref}$         & The relevance between articles.                                                          & 0.0174      \\ \cline{2-4}
		\multicolumn{1}{|c|}{}                                   & $Sim(p_{1},p_{2})$  & The cosine similarity between articles.                                                  & 0.1437      \\ \hline
		\multirow{3}{*}{Venue}                          & $PR(v_{i})$         & Venues' PageRank values in the paper-venue network.                                      & 0.2146      \\ \cline{2-4}
		& $Ave_{ci}^{v_{i}}$  & The average citations of papers published in  venues.                                 & 0.1924      \\ \cline{2-4}
		& $h(v_{i})$          & The $h$-index of venues.                                                                 & 0.2081      \\ \hline
		\multirow{9}{*}{Author}                         & $h(a_{i})$   & Each scholar's $h$-index value.                                                          & 0.9782      \\ \cline{2-4}
		& $PR_{a_{i}}$        & Each scholar's PageRank value in co-author network.                                  & 0.6274      \\ \cline{2-4}
		& $AIF$               & The author impact factor.                                                                & 0.3826      \\ \cline{2-4}
		& $Q value$           & The author's Q value.                                                                    & 0.5394      \\ \cline{2-4}
		& $hmax_{co}^{a_{i}}$ & The max $h$-index value of scholar's coauthors.                                          & 0.8253      \\ \cline{2-4}
		& $Num_{co}^{a_{i}}$ & The number of scholar's coauthors.                                          & 0.426      \\ \cline{2-4}
		& $have_{co}^{a_{i}}$ & The average $h$-index value of scholar's coauthors.                                      & 0.482       \\ \cline{2-4}
		& $hlo_{co}^{a_{i}}$  & The lowest $h$-index value of scholar's coauthors.                                       & 0.275       \\ \cline{2-4}
		& $hdif_{a_{i}}$      & \tabincell{c}{The differentials between the max and the \\lowest $h$-index value of scholar's coauthors.} & 0.538       \\ \cline{2-4}
		& $Div(a_{i})$        & The diversity of coauthors.                                                              & 0.1743      \\ \hline
		\multirow{8}{*}{Institution}                    & $h_{col}$           & The $h$-index of scholars's colleague.                                                   & 0.2947      \\ \cline{2-4}
		& $Num_{pub}^{col}$   & The number of publications of scholars's colleague.                                      & 0.1368      \\ \cline{2-4}
		& $Cits_{col}$        & The citation counts of scholars's colleague.                                             & 0.1937      \\ \cline{2-4}
		& $PR_{col}$          & The PageRank score of scholars's colleague.                                              & 0.0264      \\ \cline{2-4}
		& $G(i)^{h}$          & The Gini coefficient on $h$-index of institution.                                        & 0.0937      \\ \cline{2-4}
		& $G(i)^{Cit}$        & The Gini coefficient on citation counts of institution.                                  & 0.0153      \\ \cline{2-4}
		
		& $G(i)^{pub}$        &
		\tabincell{c}{The Gini coefficient on number of publications \\of institution.}
		& 0.1632      \\ \cline{2-4}
		& $GDP$               & The GDP value of the institution's country.                                              & 0.1937      \\ \hline
		\multirow{2}{*}{Temporal}                       & $Num_{years}$       & Scholar's academic ages.                                                                 & 0.5863      \\ \cline{2-4}
		& $Hindex$-$dif$      &
		\tabincell{c}{The difference between scholar's $h$-index and\\ $\triangle t$ years ago.               }
		& 0.6248      \\ \hline
	\end{tabular}
\end{table*}

\begin{figure*}
\centering
\includegraphics[width=13cm]{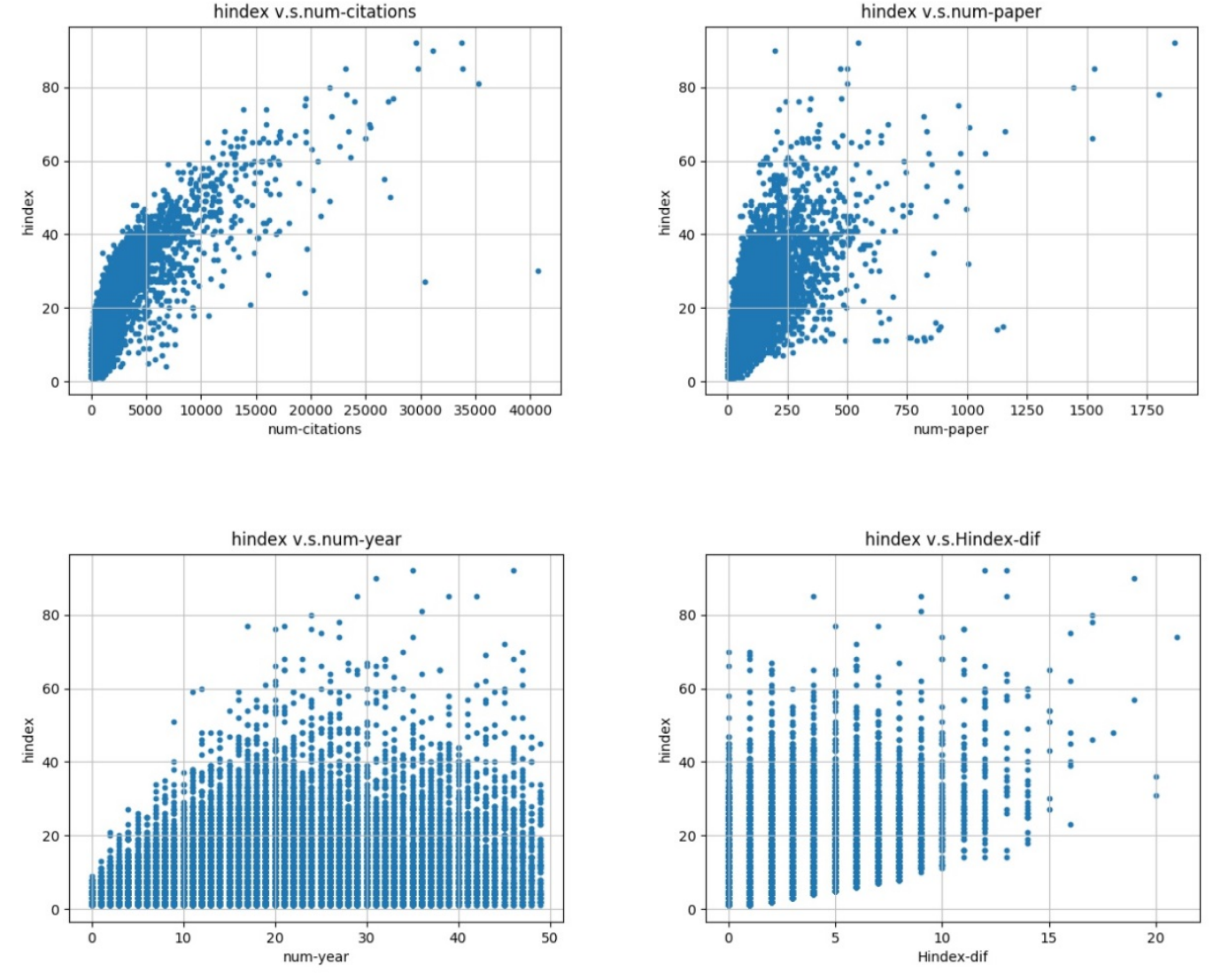}

\caption{The linear regression phenomena between the $h$-index and the four most relevant factors.}
\label{hindex}
\end{figure*}

According to Table \ref{PCC}, we can see that the author-centered and article-centered factors are the most correlative factors among all the proposed indicators following by the temporal-centered factors. To further present the linear correlation phenomenon between the $h$-index and the other factors, we then depict the results of the four most relevant factors. As shown in Figure \ref{hindex}, scholars' number of citations and publications, academic years, and the differentials of $h$-index in $\triangle t$ years are highly correlate with scholars' future $h$-index. With the increase of academic age, h-index keeps increase until academic age is 15, after that keeps steady.

It is not difficult to understand the high relevance of $Cites$, $Num_{pub}$, and $h_{a_{i}}$. While the venue-centered and institution-centered factors seem to be negatively correlated with scholar's $h$-index. However, this table cannot accurately depict the effectiveness of these factors on predicting the future academic success of scholars since the factors may reveal the same phenomenon together. From this table, it can only show the linear correlation between them, and their performance in predicting scholar's $h$-index are investigated in the next section.

\begin{figure}
    \centering
    \includegraphics[width=13cm, height=4cm]{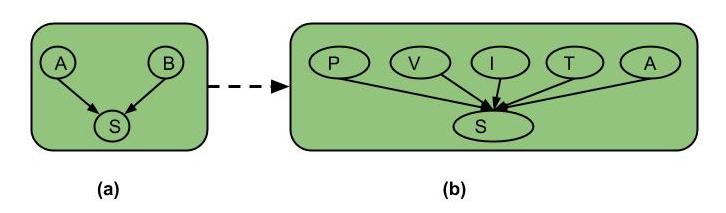}
    \caption{V-structure causal inference model}
    \label{fig:vstructue}
\end{figure}

\section{Causal Factor Verification}

\begin{figure}
\centering
\includegraphics[width=10cm]{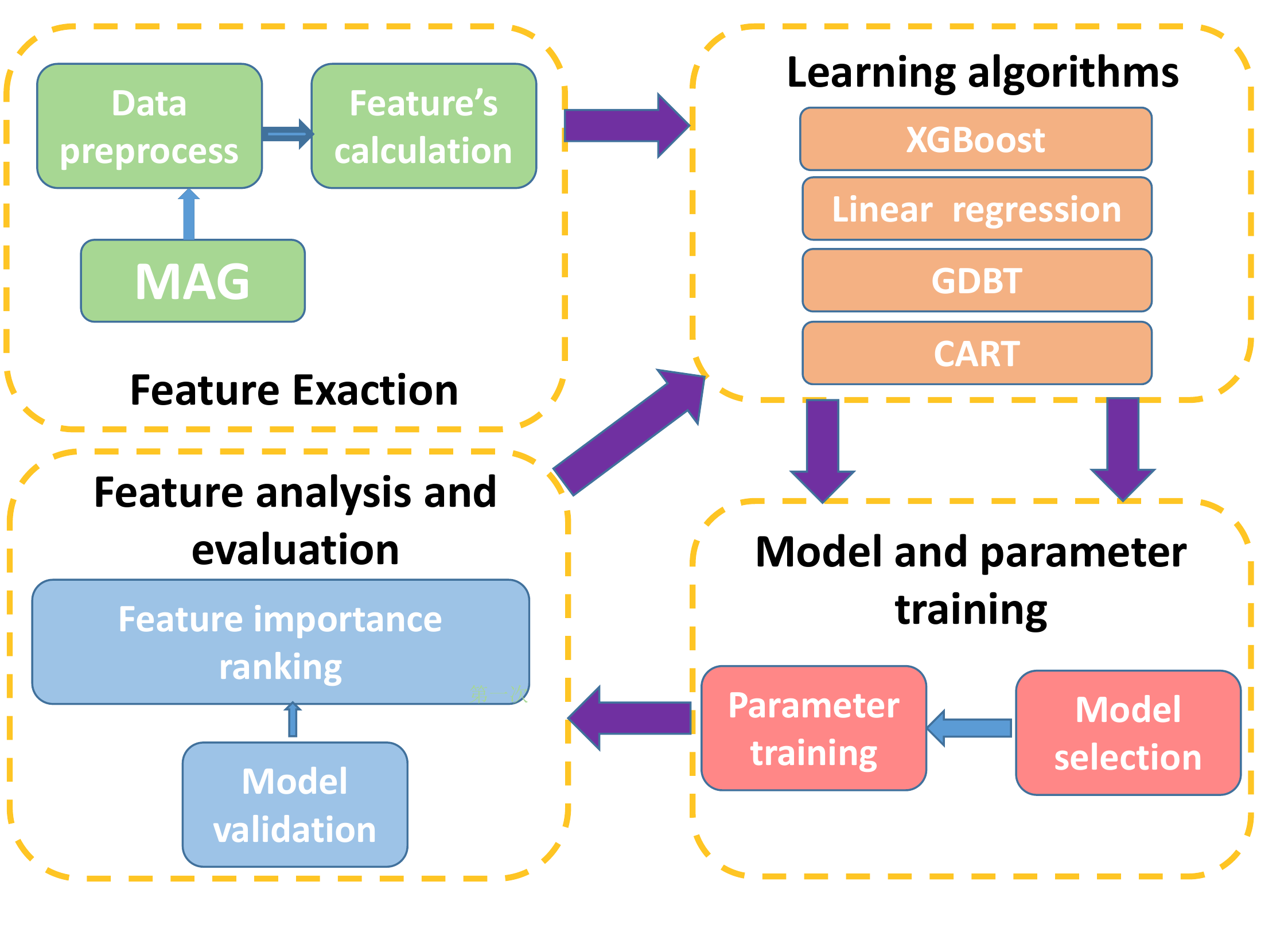}

\caption{Causal inference framework}
\label{archi}
\end{figure}

In this section, we explore the performance of the above-mentioned factors on predicting scholar's $h$-index. In order to investigate their effectiveness, we use advanced machine learning techniques. Among them, we apply the XGboost, linear regression, gradient boosting decision trees, and classification and regression tree separately on the dataset. Then, by comparing the performance, we find the most appropriate machine learning method.
\subsection{Structured Causal Model (SCM) Construction}
The core of the structural theory of causation lies a "structural causation model (SCM) \citep{Shiffrin7308,li2016observational, yang2018learning}". Therefore, in our paper, we first present a simple causal model as shown in Fig.~\ref{fig:vstructue}. Here, $S$ is a collider: arrows 'collide' at S. the path  $A\rightarrow S \leftarrow B$ is blocked. In other words, $A$ is not associated with $B$ through $S$. Given a collider $S$. the causal factors are independent with each other. We call this structure a $V$-structure \citep{le2013inferring}. In this $V$-structure, $A$ and $B$ represent parents of $S$. However, in our paper, scientific success $S$ can be determined and influenced by multiple causal factors. Therefore, we expand in Fig.~\ref{fig:vstructue}(a) as a heterogeneous structured causal model as shown in ~\ref{fig:vstructue}(b) formulated as the following equation.
\begin{tcolorbox}
\begin{align}
P(S) &= P(S|F_{1})\dots P(S|F_{n})  & \text{Causal Factor Discovery}
\end{align}
\end{tcolorbox}

\subsection{Causal Effects Prediction}

A SCM model $S$, consisting of two sets of variables, $X$ and $Y$ ,and a set $F$ of functions that determine how values are assigned to each variable $X_{i} \in X$. Here, we assume that, given the prediction output $Y$, the function $f$ represents the effect $Y$ as a function of the direct causes $X$ and marginal loss $\epsilon$ with learning parameters $\theta_{1}$. After we identify factors that contribute to our scientific success, we need to measure the significance for each causal factor. From the observational big  scholarly dataset, we apply advanced machine learning techniques to discover the causal relationships and how various causal factors, including \textit{Author, Paper, Venue, Institution and Temporal} facilitates understanding the scientific success.
\begin{tcolorbox}
\begin{align}
S &= F(X, Y, \epsilon ; \theta_{1})  & \text{Loss Function }
\end{align}
\end{tcolorbox}
\textcolor{red}{\subsubsection{Causal Factor Learning Algorithms}}

In this section, we apply the following four advanced machine learning techniques to estimate the causal effects.

\textbf{XGBoost}: XGBoost is a scalable end-to-end tree boosting
system and is faster than the most current widely used methods. The idea of the boosting algorithm is to integrate many weak classifiers together to form a strong classifier, and XGBoost is a lifting tree model which integrates many CART regression tree models to form a strong classifier. Its tree boosting mainly consists two parts, which are the regularized learning objective and the gradient tree boosting process.

\textbf{Linear Regression (LR)}: Regression analysis is widely used for prediction and forecasting, and it can also be used to find out which among all independent variables are related with the dependent variable. Linear regression requires that the model is linear in regression parameters. The predictor function is utilized to model the data, and the data can be used to estimate the unknown parameters. Linear Regression is fast in modeling and runs fast in the case of large amounts of data.

\textbf{Gradient Boosting Decision Trees (GBDT)}: GBDT is an iterative decision tree algorithm, which includes many decision trees and the final result equals to the sum of all the trees' decisions. The core of GBDT is that every tree learns the residual of the sum of all previous tree conclusions, which is the sum of the real values after adding the predicted values. It can discover a variety of distinct features and their combinations.

\textbf{Classification and Regression Trees (CART)}: It can be used to create a classification tree or a regression tree. When CART is used as a classification tree, the feature attributes can be continuous or discrete, and a CART classification tree uses Gini index in node splitting. When CART is used as regression tree, observation attributes are required to be continuous type. Because the least absolute deviation (LAD) or least square deviation (LSD) method is usually used when selecting feature attributes by node splitting, the feature attributes are also continuous type. In our paper, we apply it as a regression tree to predict scholar's future impact based on the input variables.

\subsection{Dataset}
In this paper, we use two datasets of different disciplines. One is the sub-dataset extracted from the Microsoft Academic Graph (MAG). The MAG dataset contains detailed paper information including \textit{title}, \textit{keywords}, \textit{authors}, \textit{institutions}, \textit{venues}, \textit{publication date}, and \textit{citations} from 27 macro-areas and 306 sub-areas. The whole dataset includes over 35 million papers, 38 million authors, and more than 324 million citation relationships. We use a sub-dataset includes 79,321 scholar profiles and 105,123 articles focusing on computer science domain with complete academic careers.

The other is a subset of American Physical Society (APS). The APS dataset contains physics paper information of \textit{title}, \textit{authors}, \textit{institutions}, \textit{venues}, \textit{publication date}, and \textit{citations}. The whole dataset includes 540,232 papers, 394,801 authors, and more than 6 million citation relationships of 12 APS journals. We use a sub-dataset including PRC and PRE papers, 80,360 scholar profiles and 98,011 articles in total.

\subsection{Evaluation Metrics}
In order to evaluate the performance of different learning algorithms and factors, we adopt four typical metrics including MAE (Mean Absolute Error), MAPE (Mean Absolute Percentage Error), MSE (Mean Squared Error), ACC (Accuracy), and $R^{2}$. Given the true value $y$, and the predictive value $\hat{y}$, the above-mentioned evaluation metrics can be calculated as follows:
\begin{equation}
\label{equ:MAE}
MAE = \frac{1}{n}\sum_{i=1}^{n}|\hat{y}-y|
\end{equation}

\begin{equation}
\label{equ:MAPE}
MAPE = \frac{100}{n}\sum_{i=1}^{n}|\frac{y-\hat{y}}{y}|
\end{equation}

\begin{equation}
\label{equ:MSE}
MSE = \frac{1}{n}\sum_{i=1}^{n}|\hat{y}-y|^{2}
\end{equation}

\begin{equation}
\label{equ:}
ACC = \frac{1}{n}\sum_{i=1}^{n}I(f(y_{i})=\hat{y})
\end{equation}

\begin{equation}
\label{equ:R}
R^{2} = 1-\frac{\sum_{i=1}^{n}|y-\hat{y}|^{2}}{\sum_{i=1}^{n}|y-\bar{y}|^{2}}
\end{equation}

\subsection{Validation by experiments}
With the learning algorithms and factors we introduced above, scholar's $h$-index can be predicted. We use the previous $\triangle t$ years' information for training, and the real data in 2015 (MAG) or 2013 (APS) to validate. On the train set, we perform a 5-folds cross-validation to tune the hyperparameter of models. For XGBoost and GBDT, the main hyperparameters we have tuned include the learning rate, maximum depth of trees, sub-sample rate, sub-feature rate and the regularization coefficient. For CART, the main hyperparameters we have tuned include the learning rate, maximum depth of trees. For LR, the main hyper parameters we have tuned include the learning rate and the regularization coefficient. All hyper parameters are tuned by grid search on the parameter space. The results are illustrated from the aspect of MAE, MAPE, MSE, ACC, and $R^{2}$.

Table \ref{ACC} shows the predictive performances of different methods on the evaluation metrics mentioned above on MAG dataset. MSE, MAE, and MAPE are used to compare the predictive results and the true values. In the table, $R^{2}$ indicates the correlation between the predictive results and the true values and $ACC$ indicates the accuracy. Hence, the better prediction performance can be inferred by their values. It is obvious that the performance of XGBoost is the best among all the methods using different time periods because it outperforms other methods on 4 of 5 metrics, which gets the smallest MAPE and MES and highest Acc and $R^2$ score in three groups experiments. While for different $\triangle t$ values, there exist various prediction results. The performance of $\triangle t$=7 achieves the best score, and the results by $\triangle t$=10 are the worst. However, there only exists a slight difference between the results of $\triangle t$=5 and $\triangle t$=7. In the following parts, we analyze the results in $\triangle t$=7 on the MAG dataset.

\begin{table}[htbp]
\centering
\caption{The Performance of Difference Learning Algorithms on MAG.}
\label{ACC}
\begin{tabular}{lcccccc}
\hline
                                  & \multicolumn{1}{l}{} & MAE                      & MAPE                     & MSE                      & ACC                      & $R^{2}$ \\ \hline
\multirow{4}{*}{$\triangle t$=5}  & XGBoost              & 0.73                     & 0.07                     & 1.09                     & 0.86                     & 0.99   \\
                                  & LR                   & 0.82                     & 0.10                     & 1.18                     & 0.80                     & 0.92   \\
                                  & GBDT                 & 0.69                     & 0.08                     & 1.16                     & 0.84                     & 0.95   \\
                                  & CART                 & 0.96                     & 0.18                     & 2.30                     & 0.79                     & 0.81   \\ \hline
\multirow{4}{*}{$\triangle t$=7}  & XGBoost              & 0.79                     & 0.07                     & 1.19                     & 0.86                     & 0.99   \\
                                  & LR                   & 0.83                     & 0.11                     & 1.30                     & 0.79                     & 0.90   \\
                                  & GBDT                 & 0.73                     & 0.09                     & 1.25                     & 0.85                     & 0.94   \\
                                  & CART                 & 0.98                     & 0.20                     & 2.49                     & 0.78                     & 0.80   \\ \hline
\multirow{4}{*}{$\triangle t$=10} & XGBoost              & 0.81                     & 0.09                     & 1.27                     & 0.83                     & 0.91   \\
                                  & LR                   & 0.84                     & 0.13                     & 1.43                     & 0.73                     & 0.86   \\
                                  & GBDT                 & 0.74                     & 0.10                     & 1.32                     & 0.81                     & 0.84   \\
                                  & CART                 & 0.99                     & 0.29                     & 2.68                     & 0.74                     & 0.63   \\ \hline
\end{tabular}
\end{table}

The predictive performances of these methods on the APS dataset are shown in Table \ref{ACC_APS}. In the table, we observed that the overall prediction performances on APS are better than on MAG, which has smaller errors and higher fitting degree ($R^2$).
We noticed that the performance of XGBoost is also the best. Different from the results on MAG, all methods perform better on the period $\triangle t = 10$ than other groups. In the following parts, we analyze the results in $\triangle t$=10 on the APS dataset.

\begin{table}[htbp]
\centering
\caption{The Performance of Difference Learning Algorithms on APS.}
\label{ACC_APS}
\begin{tabular}{lcccccc}
\hline
                                  & \multicolumn{1}{l}{} & MAE                      & MAPE                     & MSE                      & ACC                      & $R^{2}$ \\ \hline
\multirow{4}{*}{$\triangle t$=5}  & XGBoost              & 0.54                     & 0.06                     & 0.62                     & 0.81                     & 0.97   \\
                                  & LR                   & 0.57                     & 0.08                     & 0.65                     & 0.80                     & 0.96   \\
                                  & GBDT                 & 0.55                     & 0.06                     & 0.63                     & 0.80                     & 0.96   \\
                                  & CART                 & 0.56                     & 0.09                     & 0.73                     & 0.78                     & 0.95   \\ \hline
\multirow{4}{*}{$\triangle t$=7}  & XGBoost              & 0.51                     & 0.07                     & 0.55                     & 0.82                     & 0.97   \\
                                  & LR                   & 0.53                     & 0.10                     & 0.55                     & 0.80                     & 0.96   \\
                                  & GBDT                 & 0.51                     & 0.07                     & 0.56                     & 0.81                     & 0.97   \\
                                  & CART                 & 0.54                     & 0.08                     & 0.70                     & 0.80                     & 0.96   \\ \hline
\multirow{4}{*}{$\triangle t$=10} & XGBoost              & 0.45                     & 0.06                     & 0.46                     & 0.85                     & 0.98   \\
                                  & LR                   & 0.47                     & 0.07                     & 0.48                     & 0.83                     & 0.97   \\
                                  & GBDT                 & 0.45                     & 0.08                     & 0.47                     & 0.84                     & 0.97   \\
                                  & CART                 & 0.46                     & 0.07                     & 0.60                     & 0.80                     & 0.96   \\ \hline
\end{tabular}
\end{table}

\subsection{Causal Factor Evaluation}
The above analyses verify the causal relationships between various factors and the overall $h$-index results. However, the contribution and importance of factors still need to be explored. To solve this question, we first apply the "jackknife" method \citep{severiano2011evaluation} to verify the function of each group's factors separately. The "jackknife" method includes two phases: \textit{Adding} and \textit{Removing}. During \textit{Adding} phase, we use one group of factors each time to predict the result. During \textit{Removing} phase, we remove a group of factors and train the model with the rest factors. After these two phases, factor's individual contribution to the overall prediction task can be explored.


\begin{figure*}
\centering
\includegraphics[width=0.95\textwidth]{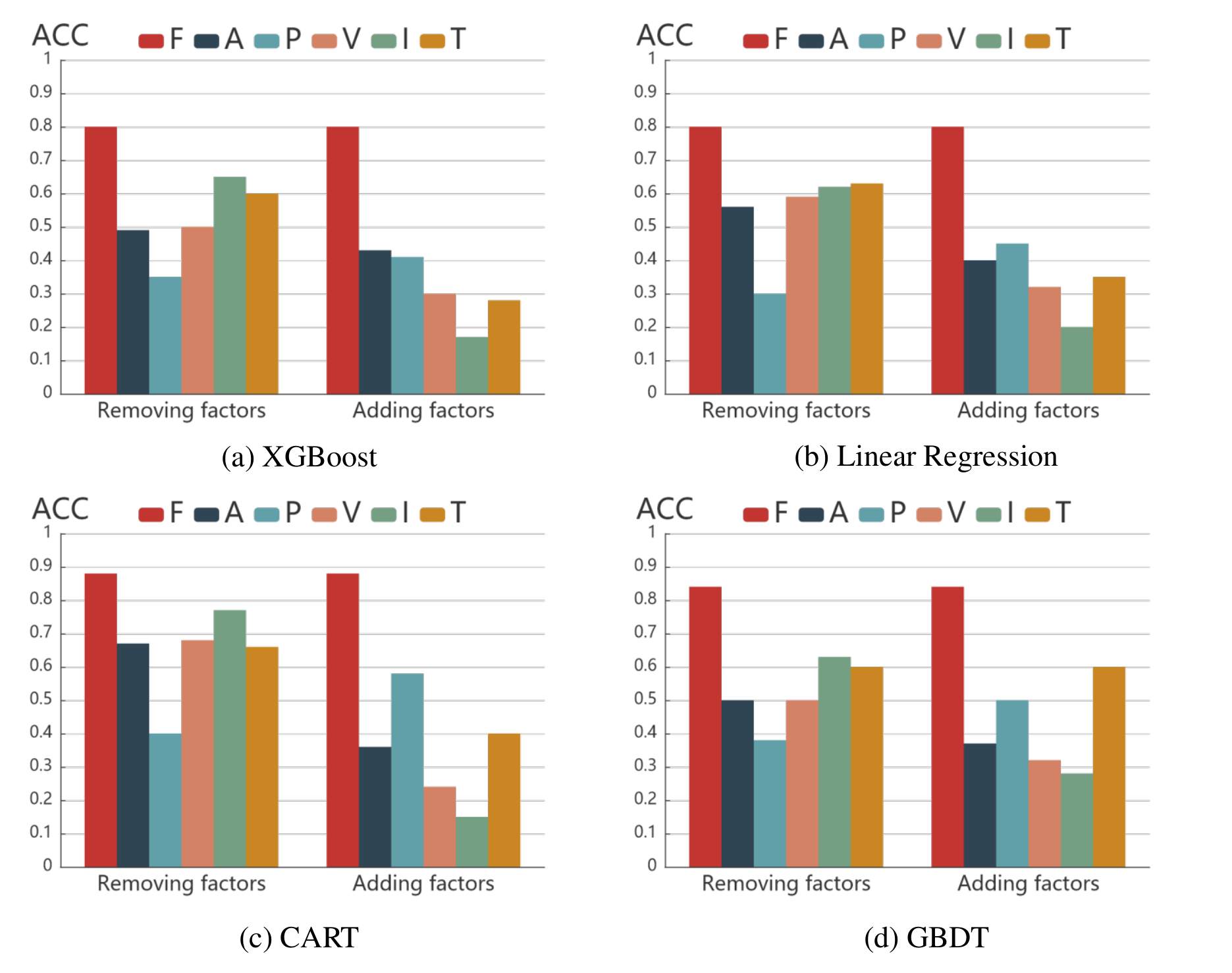}
\caption{Factor contribution analysis on MAG. Four models trained with only or without the denoted factors. F: full feature set; A: Author factors; P: Paper factors; V: Venue factors; I: Institution factors; T: Temporal factors.}
\label{jk_mag}
\end{figure*}

As shown in Figure \ref{jk_mag}, in the experiments on MAG dataset, the drop in ACC values by the removal of article-centered factors in the four methods demonstrate that they are of great significance in predicting the $h$-index. On the contrary, when removing other types of factors, the decline of the predictive performance is not so obvious. This fact can reveal the importance of article-centered factors on predicting scholars' future success. For adding factors, article-centered factors still show their important roles in predicting the future scientific impact. Moreover, author-centered and temporal-centered factors also show their effectiveness for scholar's $h$-index prediction.


\begin{figure*}
\centering
\includegraphics[width=0.95\textwidth]{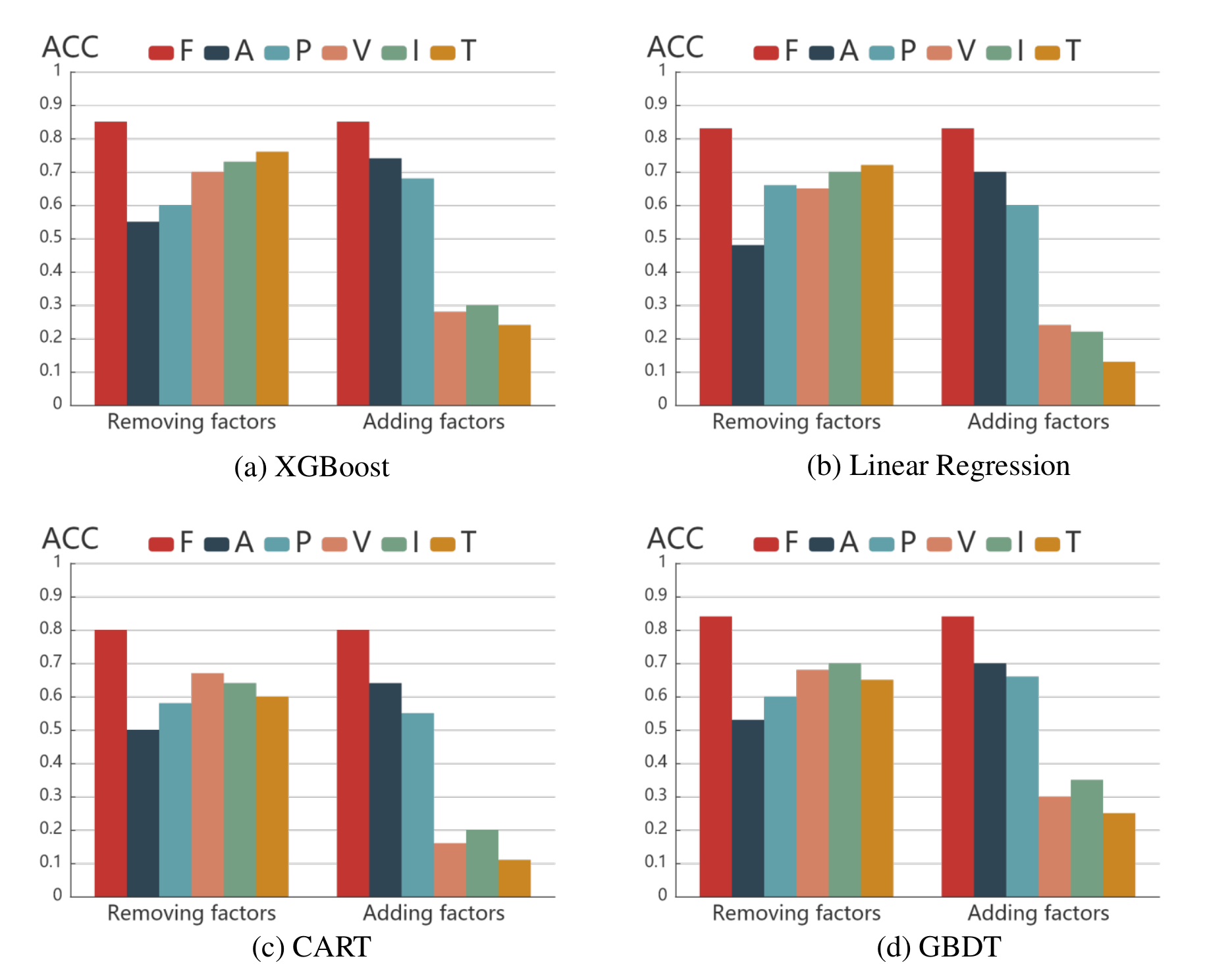}
\caption{Factor contribution analysis on APS. Four models trained with only or without the denoted factors. F: full feature set; A: Author factors; P: Paper factors; V: Venue factors; I: Institution factors; T: Temporal factors.}
\label{jk_aps}
\end{figure*}

The same analyses are performed on APS dataset, as shown in Figure \ref{jk_aps}. Different from the results on MAG, the drop in ACC values by the removal of author-centered factors greatly influences the predicting of $h$-index, which indicates that the author-centered factors are of great significance in the APS dataset. Same as previous experiments, when removing other types of factors, the decline of the predictive performance is not so obvious. For adding factors, author-centered factors still show their important roles. Moreover, article-centered factors also show their effectiveness for scholar's $h$-index prediction in APS dataset, but other features have no obvious effects, which differs from experiment results on MAG.

\begin{figure*}
\centering
\includegraphics[width=0.95\textwidth]{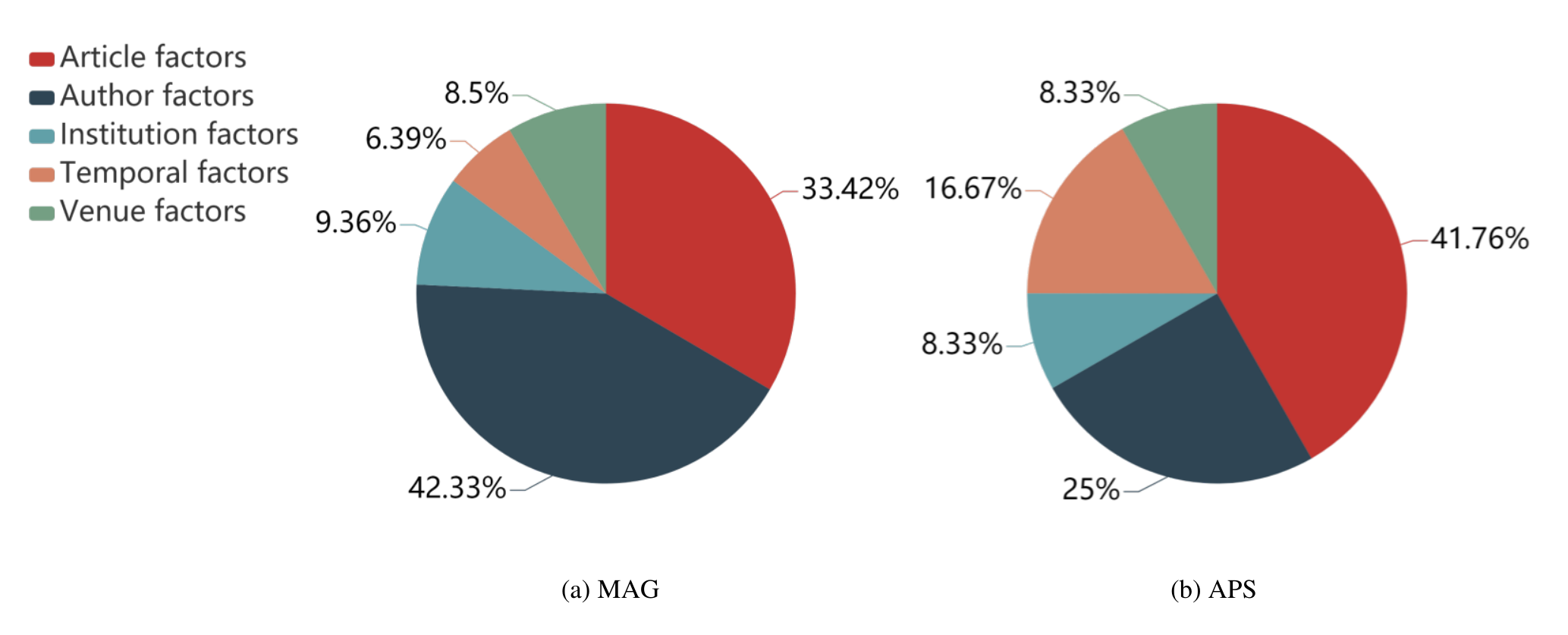}
\caption{The importance score of different factors.}
\label{factor2}
\end{figure*}

Furthermore, we analyzed the feature importance given by XGBoost of results on MAG and APS dataset. In XGBoost, feature importance can be calculated as times that a feature has been used to divide samples on leaves of trees in the model. The more frequently a feature has been used, the more important it is to the model. As shown in Figure \ref{factor2}, in both datasets, the top influential factors are still the same with the above conclusions, where the article-centered factors are still the most important for predicting the future academic success in MAG, which takes 41.67\% importance scores; and the author-centered factors are importance in APS, which takes 42.33\% importance scores.



In addition, we also analyze the Gini Coefficient of different institutions. Gini Coefficient smaller than 0.2 indicates the institutions are absolute equal. Gini Coefficient from 0.2 to 0.3 indicates the institutions are relatively equal. Gini Coefficient from 0.3 to 0.4 indicates the institutions are relatively rational. Finally, Gini Coefficient greater than 0.4 indicates the great disparity between institutions.

%

To have a comprehensive Gini coefficient of an institution, we first rank the institutions according to the number of scholars. Then according to the ranking list, the Gini coefficient on citations, number of publications, and $h$-index can be obtained. We show the Gini coefficient of top 5\%, 10\%, 20\%, and the last 10\%. As shown in Table \ref{Gini} and Table \ref{Gini_APS}, there exist some interesting phenomena. For top 5\% ranking institutions, both in MAG and APS, their Gini Coefficient values are under 0.2, which indicate that scholars' $h$-index is very close to their colleagues in the same institution. In top 10\% ranking institutions, except for citations, the Gini Coefficients for the number of papers and $h$-index are under 0.2, which still show the equality of scholars on these two aspects. While for institutions in top 20\% and last 10\%, their Gini Coefficient for the number of papers and citation exceed 0.2. It is apparent that there exist some differences in the number of papers and citation of researchers in these institutions. However, the $h$-index level of scholars in all the institutions mentioned above is very similar to their colleagues. This phenomenon shows that scholars in the same institution are birds of s feather flock together. And this phenomenon is the same in computer science and physics. The reason behind this is that the scholarly communication among them is very convenient and frequent, and they can directly feel the peer pressure from their colleagues to some extent. Therefore, scholars are trying to keep up with their colleagues in academic research, and their overall scientific impacts are quite similar to each other. Also, when providing faculty positions for researchers, there may exist standard hiring requirements for the same institution. As a consequence, the scholars in the same institution are at the same academic level.

\begin{table}[htbp]
\centering
\caption{The average Gini Coefficients of top ranking institutions on MAG.}
\label{Gini}
\begin{tabular}{c|l|l|l}
\hline
                       & Number of papers                      & Citation                      & $h$-index                      \\ \hline
Top 5\%                & 0.102418                              & 0.161101                      & 0.042667                       \\
Top 10\%               & 0.191524                              & 0.277041                      & 0.091791                       \\
Top 20\%               & 0.230564                              & 0.327122                      & 0.121142                       \\
Last 10\%              & 0.351485                              & 0.452952                      & 0.200423                       \\ \hline
\end{tabular}
\end{table}

\begin{table}[htbp]
\centering
\caption{The average Gini Coefficients of top ranking institutions on APS.}
\label{Gini_APS}
\begin{tabular}{c|l|l|l}
\hline
                       & Number of papers                      & Citation                      & $h$-index                      \\ \hline
Top 5\%                & 0.011092                              & 0.070037                      & 0.015215                       \\
Top 10\%               & 0.169559                              & 0.205179                      & 0.145778                       \\
Top 20\%               & 0.216014                              & 0.249943                      & 0.184754                       \\
Last 10\%              & 0.266891                              & 0.297307                      & 0.228714                       \\ \hline
\end{tabular}
\end{table}

\section{Conclusion and Future Work}

In this paper, we aim at discovering the causal factors that play crucial roles in predicting the scholars' academic success. To solve this issue, we first propose five potential causal factors, which are the article-centered factors, author-centered factors, venue-centered factors, institution-centered factors, and temporal factors. Then by utilizing the state of the art machine learning algorithms, we find that the article and author-centered factors are most significant causal factors for forecasting scholars' future success.

Furthermore, we analyze each factor's contribution by using the "jackknife" method and grading factors during the predicting process. The results further demonstrate the importance of article and author-centered factors.
We further analyze the specific importance ranking of these five groups of factors used in our experiments. After this process, we find that, in the MAG dataset, the article-centered factors have 41.47\% importance, the author-centered factors are in 25\% importance, the temporal-centered factors are in 16.67\% importance, and the venue and institution-centered factors are in 8.33\% importance, while in the APS dataset, the article-centered factors have 33.42\% importance, the author-centered factors are in 42.33\% importance, the temporal-centered factors are in 6.39\% importance, and the venue-centered factors have 8.50\% importance , and institution-centered factors are in 9.36\% importance. Meanwhile, we also find that the $h$-index of scholars in the same institutions tend to be very close to each other.

In the future, we plan to identify more factors and conduct our experiments on other datasets from various disciplines to demonstrate the validity of our work.

\bibliographystyle{ACM-Reference-Format}
\bibliography{Gene_Final}

\end{document}